\documentclass[fleqn]{llncs}

\input{epsf.sty}
\def\smallromani{\renewcommand{\theenumi}{\roman{enumi}}
        \renewcommand{\labelenumi}{(\theenumi)}}

\newcommand{\rrarrow}{\longrightarrow}
\newcommand{\labtran}[1]{\stackrel{#1}{\rrarrow}}

\newcommand{\indrule}[2]{\frac{\raisebox{1ex}{$#1$}}{\raisebox{-1.5ex}{$#2$}}}

\newcommand{\ls}{[\!(}
\newcommand{\rs}{)\!]}
\newcommand{\nooutpref}{\alpha}
\newcommand{\catarrow}[2]{\mathrel{\mathop{\kern 0pt \rrarrow} \limits_{#1}^{#2}}}
\newcommand{\prob}{{\it pb}}

\long\def\comment#1{}

\begin{document}

\title{Probabilistic asynchronous $\pi$-calculus}
\author{Oltea Mihaela Herescu and Catuscia Palamidessi}
\institute{Dept. of Comp. Sci. and Eng.,
The Pennsylvania State University \\
University Park, PA 16802-6106  USA
\\ 
{\tt \{herescu,catuscia\}@cse.psu.edu} }
\date{}
\maketitle

\begin{abstract}
We propose an extension of the asynchronous 
$\pi$-calculus with a notion of random choice. 
We define an operational semantics which distinguishes 
between probabilistic choice, made internally by the process, 
and nondeterministic choice, made externally by an adversary 
scheduler. This distinction will allow us to reason about the 
probabilistic correctness of algorithms under certain schedulers. 
We show that in this language we can solve the electoral problem, 
which was proved not possible in the asynchronous $\pi$-calculus.
Finally, we show an implementation of the probabilistic 
asynchronous $\pi$-calculus
in a Java-like language.
\end{abstract}

\section{Introduction}
The $\pi$-calculus (\cite{Milner:92:IC,Milner:93:TCS}) 
is a very expressive specification language 
for concurrent programming, but the difficulties in its distributed 
implementation challenge its candidature to be a canonical model of 
distributed computation. Certain mechanisms of the $\pi$-calculus, in fact, 
require solving a problem of distributed consensus. 

The asynchronous $\pi$-calculus (\cite{Honda:91:ECOOP,Boudol:92:REPORT}), 
on the other hand, is more suitable for a distributed implementation, 
but it is rather weak for solving distributed problems
(\cite{Palamidessi:97:POPL}). 
 
In order to increase the 
expressive power of the asynchronous $\pi$-calculus
we propose a probabilistic extension, 
$\pi_{pa}$, based on the probabilistic automata of Segala and Lynch 
(\cite{Segala:95:NJC}). The characteristic of this model is that 
it distinguishes between probabilistic and nondeterministic behavior. 
The first is associated with the random choices of the process, while 
the second is related to the arbitrary decisions of an external scheduler. 
This separation allows us to reason about adverse conditions, i.e. schedulers 
that ``try to prevent'' the process from achieving its goal.
Similar models were presented in \cite{Vardi:85:FOCS} and \cite{Yi:92:IFIP}.

Next we show an example of distributed problem that can be solved with 
$\pi_{pa}$, namely the election of a leader in 
a symmetric network. It was proved in \cite{Palamidessi:97:POPL} that 
such problem cannot be solved with the 
asynchronous $\pi$-calculus. We propose an algorithm for the solution 
of this problem, and we prove that it is correct, i.e. that the leader 
will eventually be elected, with probability $1$, under every possible 
scheduler. Our algorithm is reminiscent of the  algorithm used in
\cite{Rabin:94:HOARE} for solving the dining philosophers problem, 
but in our case we do not need the fairness assumption. Also, the fact that 
we give the solution in a language provided with a rigorous 
operational semantics
allows us to give a more formal proof of correctness.

Finally, we define a ``toy'' distributed 
implementation of the $\pi_{pa}$-calculus
into a Java-like language. The purpose of this exercise is 
to prove that $\pi_{pa}$ is a reasonable paradigm for the
specification of distributed algorithms, since it can be 
implemented without loss of expressivity.

The novelty of our proposal, 
with respect to other probabilistic process algebras
which have been defined in literature 
(see, for instance, \cite{Glabbeek:95:IC}), 
is the definition of the parallel operator 
in a CCS  style, as opposed to the  SCCS 
style. Namely, parallel processes 
are not forced to proceed simultaneously. 
Note also that for general 
probabilistic automata it is not possible 
to define the parallel operator (\cite{Segala:95:PhD}), 
or at least, there is no natural definition. 
In $\pi_{pa}$ 
the parallel operator can be defined as a natural extension of the 
non probabilistic case, and this 
can be considered, to our opinion, 
another argument in favor of the suitability of $\pi_{pa}$  
for distributed implementation.

\section{Preliminaries} 
In this section we recall 
the definition of the asynchronous $\pi$-calculus
and the definition of probabilistic automata. 
We consider the {\it late} semantics of the $\pi$-calculus, 
because the probabilistic extension of the late 
semantics  is simpler
than the eager version. 

\subsection{The asynchronous $\pi$-calculus}
We follow the definition of the 
asynchronous $\pi$-calculus given in \cite{Amadio:98:TCS}, 
except that  we will use recursion instead of the 
replication operator, since we find it to be more convenient 
for writing programs. It is well known that 
recursion and replication
are equivalent, 
see for instance \cite{Milner:99:BOOK}. 

Consider a countable set of {\it channel names}, 
$x, y,\ldots$, and a countable set of {\it process names}
$X, Y,\ldots$. The  prefixes   
$\alpha, \beta,\ldots$
and the processes $P,Q,\ldots$ of the
asynchronous $\pi$-calculus 
are defined by the following 
grammar: 

\[
\begin{array}{rlcl}
\mbox{\it Prefixes}&\nooutpref&\mbox{::=}&x(y)\;\;|\;\; \tau \\
{\it Processes}&P&\mbox{::=}&\bar{x}y \;\;|\;\; 
     \sum_i\nooutpref_i.P_i\;\;|\;\; 
     \nu x P \;\;|\;\; P\;|\;P \;\;|\;\; X \;\;|\;\; {\it rec}_XP
\end{array}
\]

The basic actions are 
$x(y)$, which represents the {\it input} of the (formal) name $y$ 
from channel $x$, 
$\bar{x}y$, which represents the {\it output} of the name $y$ 
on channel $x$, and
$\tau$, which stands for any silent (non-communication) action. 

The process $\sum_i\alpha_i.P_i$ represents guarded choice 
on input or silent prefixes, 
and it is usually assumed to be finite. 
We will use the abbreviations 
${\bf 0}$ ({\it inaction}) to represent the empty sum, 
$\alpha.P$ ({\it prefix}) to represent sum on one element only, and 
$P+Q$ for the binary sum. 
The symbols $\nu x$ and $|$ are the {\it restriction} and 
the {\it parallel} operator, respectively.
We adopt the convention that the prefix operator has priority wrt $+$ and $|$.
 The process
${\it rec}_XP$ represents a process $X$ defined 
as $X \stackrel{\rm def}{=} P$, where $P$ may contain occurrences of $X$
(recursive definition).  We assume that all the occurrences of $X$ in $P$ 
are prefixed. 

The operators $\nu x$ and  $y(x)$ are $x$-{\it binders}, 
i.e. in the processes  $\nu x P$ and $y(x).P$ the occurrences 
of $x$ in $P$ are considered {\it bound}, 
with the usual rules of scoping. 
The {\it free names} of $P$, i.e. those names which do 
not occur in the scope of any binder, 
are denoted by  ${\it fn}(P)$. 
The {\it alpha-conversion} of bound names is defined as usual, 
and the renaming (or substitution) $P[y/x]$ is defined as the 
result of replacing all free occurrences of $x$ in $P$ by $y$, possibly 
applying alpha-conversion in order to avoid capture. 

The operational semantics is specified via a transition system 
labeled by {\it actions} $\mu, \mu'\ldots$. 
These are given by the following grammar: 
\[
\begin{array}{rlcl}
{\it Actions} &\mu &\mbox{::=}& x(y) \;\;|\;\; \bar{x}y \;\;|\;\; \bar{x}(y)
   \;\;|\;\; \tau
\end{array}
\]
Essentially, we have all the actions from the syntax, 
plus the {\it bound output} $\bar{x}(y)$. This is introduced to model 
{\it scope extrusion}, i.e. the result of sending to another process 
a private ($\nu$-bound) name. 
The bound names of an action $\mu$, ${\it bn}(\mu)$,
are defined as follows:
${\it bn}(x(y))={\it bn}(\bar{x}(y))=\{y\}$; 
${\it bn}(\bar{x}y)={\it bn}(\tau)=\emptyset$.
Furthermore, we will indicate by $n(\mu)$ all the 
{\it names} which occur in $\mu$.

The rules for the late semantics 
are given in Table~\ref{TS}. 
The symbol $\equiv$ used in {\sc Cong}
stands for {\it structural congruence}, 
a form of equivalence which identifies 
``statically'' two processes
and which is used to simplify the presentation. 
We assume this congruence to satisfy the following: 
\begin{enumerate}
\smallromani
\item $P\equiv Q$ if $Q$ can be 
obtained from $P$ by alpha-renaming, notation $P\equiv_\alpha Q$,
\item $P\;|\;Q\equiv Q\;|\;P$,
\item ${\it rec}_XP \equiv P[{\it rec}_XP/X]$,
\end{enumerate}

\begin{table}[ht] 
\begin{center}
\begin{tabular}{|@{\ \ \ }ll@{\ \ }|}  
\hline
\mbox{}&\mbox{}
\\
{\sc Sum}
  &$\sum_i\alpha_i.P_i \labtran{\alpha_j} P_j$
\\
&\mbox{}
\\
{\sc Out}
  &$\bar{x}y \labtran{\bar{x}y}{\bf 0}$
\\
&\mbox{   }
\\
{\sc Open}
 &$\indrule{P\labtran{\bar{x}y}P'}{\nu y P\labtran{\bar{x}(y)} P'}$
  \ \ \ \ $x\neq y$
\\
&\mbox{}
\\
{\sc Res}
 &$\indrule{P\labtran{\mu}P'}{\nu y P\labtran{\mu}\nu y P'}$
  \ \ \ \ $y\not\in n(\mu)$
\\
&\mbox{}
\\
{\sc Par}
 &$\indrule{P\labtran{\mu}P'}{P \;|\; Q\labtran{\mu}  P' \;|\; Q}$
  \ \ \ \ ${\it bn}(\mu)\cap{\it fn}(Q) =\emptyset$
\\
&\mbox{}
\\
{\sc Com}
 &$\indrule{P\labtran{\bar{x}y}P'\ \ \ \ Q\labtran{x(z)}Q'}
 {P \;|\; Q\labtran{\tau} P' \;|\; Q'[y/z]}$
\\
&\mbox{}
\\
{\sc Close}
 &$\indrule{P\labtran{\bar{x}(y)}P'\ \ \ \ Q\labtran{x(y)}Q'}
  {P \;|\; Q\labtran{\tau} \nu y(P' \;|\; Q')}$
\\
&\mbox{}
\\
{\sc Cong}
 &$\indrule{P\equiv P' \ \ \ \ P'\labtran{\mu}Q' \ \ \ \ Q'\equiv Q}
  {P\labtran{\mu}Q}$
\\
&\mbox{} 
\\
\hline
\end{tabular}
\caption{The late-instantiation transition system of the asynchronous 
$\pi$-calculus.}
\label{TS}
\end{center}
\end{table}

Note that communication is modeled 
by handshaking (Rules {\sc Com} and {\sc Close}).
The reason why this calculus is considered a paradigm 
for {\it asynchornous}
communication is that there is no primitive {\it output prefix}, 
hence no primitive 
notion of continuation after the execution of an output action. 
In other words, 
the process executing an ouptut action will not be able to detect 
(in principle) when the 
corresponding input action is actually executed.

\subsection{Probabilistic automata, adversaries,
 and executions}\label{automata}
Asynchronous automata have been proposed in \cite{Segala:95:NJC}. 
We simplify here the original definition, 
and tailor it to what we need for defining the probabilistic extension of the
 asynchronous $\pi$-calculus. 
The main difference is that we consider only discrete probabilistic
spaces, and that the concept of 
deadlock is simply a node with no out-transitions. 

\begin{figure}[ht]
\begin{center}
\epsfbox{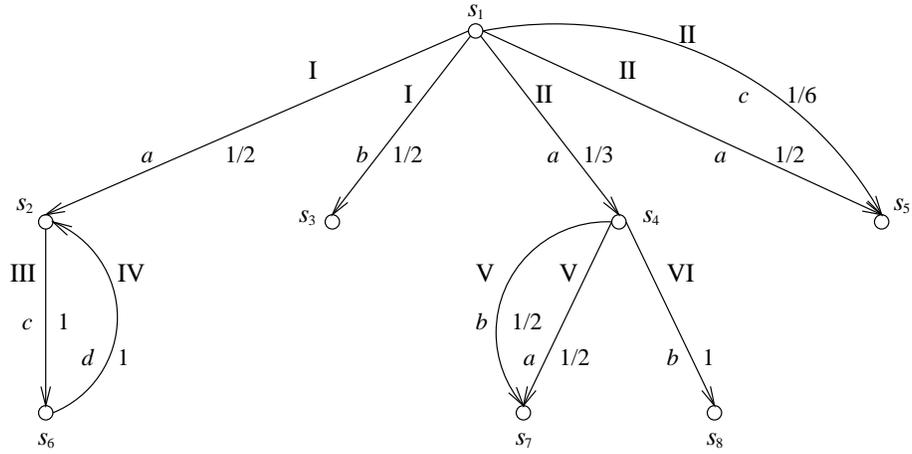} 
\caption{Example of a probabilistic automaton $M$.
The transition groups are labeled by I, II, ..., VI}
\label{fig:PA}
\end{center}
\end{figure}

\begin{figure}[ht]
\begin{center}
\epsfbox{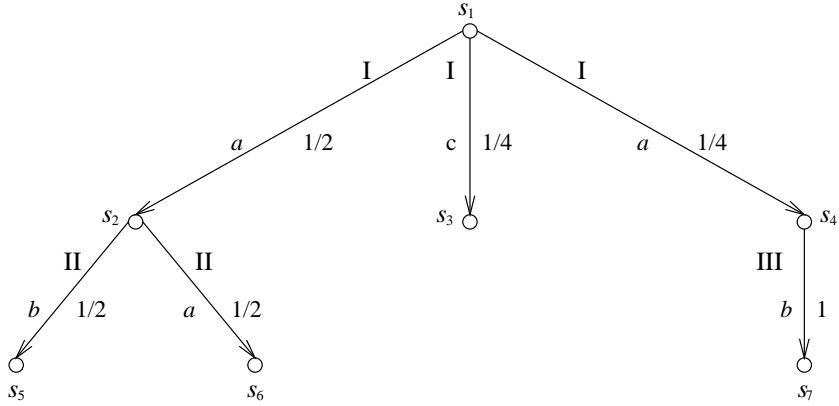} 
\caption{A fully probabilistic automaton\label{fig:FPA}}
\end{center}
\end{figure}

A discrete probabilistic space is a pair $(X,\prob)$ 
where  $X$ is a set and $\prob$ is a function 
$\prob: X \rightarrow (0,1]$ such that $\sum_{x\in X}\prob(x) = 1$.
Given a set $Y$, we define
\[
{\it Prob}(Y) = 
\{(X,\prob) \;|\; X \subseteq Y \mbox{ and } (X,\prob) 
\mbox{ is a discrete probabilistic space} \}.
\]
Given a set of states $S$ and a set of actions $A$, 
a {\em probabilistic automaton} on $S$ and $A$ is a triple 
$(S,{\cal T},s_0)$ where 
$s_0\in S$ (initial state) and ${\cal T} \subseteq S \times 
{\it Prob}(A\times S)$. 
We call the elements of ${\cal T}$ {\it transition groups}
(in \cite{Segala:95:NJC} they are called {\it steps}). 
The idea behind this model is that the choice 
between two different groups is made nondeterministically and possibly 
controlled by an external agent, e.g. a scheduler, while the transition 
within the same group is chosen probabilistically and it is controlled
internally (e.g. by a probabilistic choice operator). 
If at most  one transition group is allowed for each state, the automaton is
called {\it fully probabilistic}.
Figures~\ref{fig:PA} and \ref{fig:FPA} give examples of a probabilistic and a 
fully probabilistic automaton, respectively.

In \cite{Segala:95:NJC} it is remarked that this notion of automaton 
subsumes and extends both the the {\em reactive} and {\em generative}
models of probabilistic processes (\cite{Glabbeek:95:IC}). In particular,
the generative model corresponds to the notion of fully probabilistic 
automaton. 

We define now the notion of execution of an automaton
under a {\it scheduler}, by adapting and simplifying the 
corresponding notion given in \cite{Segala:95:NJC}.
A scheduler can be seen as a function which 
solves the nondeterminism of the automaton by
selecting, at each moment of the computation, 
a transition group among all the ones
allowed in the present state. 
Schedulers are sometimes called {\it adversaries}, 
thus conveying the idea of an external entity 
playing ``against'' the process. A process is {\it robust} wrt 
a certain class of adversaries if it gives the 
intended result for each possible scheduling 
imposed by an adversary in the class. 
Clearly, the reliability of an algorithm depends 
on how ``smart'' the adversaries 
of this class can be.
We will assume that an adversary can decide the next 
transition group depending not only on the current state, 
but also on the whole history of the computation till that moment,
including  the random choices made by the automaton. 

\begin{figure}[ht]
\begin{center}
\epsfbox{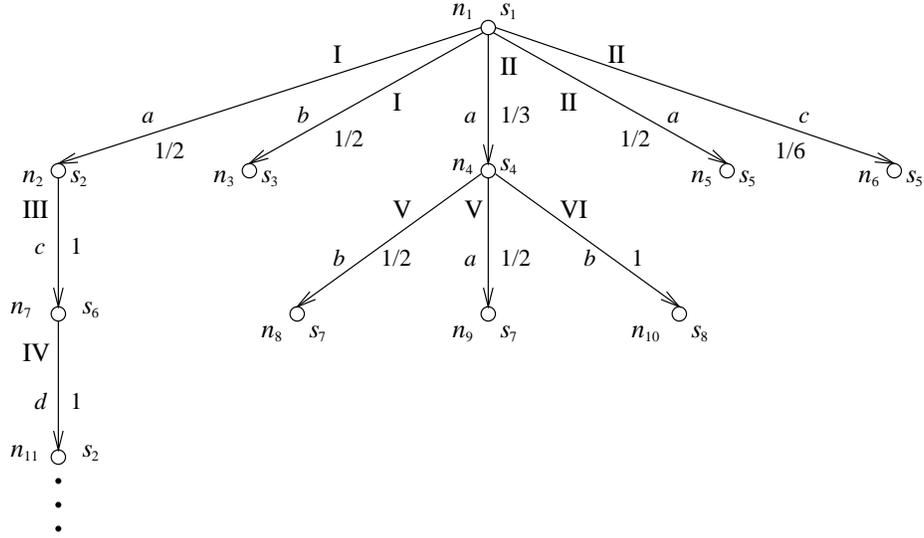}
\caption{${\it tree}(M)$, where $M$ is the  
probabilistic automaton $M$ of Figure~\ref{fig:PA}}
 \label{fig:Tree}
\end{center}
\end{figure}

Given a probabilistic automaton $M = (S,{\cal T},s_0)$, define ${\it tree}(M)$
as the tree
obtained by unfolding the transition system, 
i.e. the tree with a root $n_0$ labeled by $s_0$, and 
such that, for each node $n$, if $s\in S$ is the label of $n$, then for each
$(s, (X, \prob)) \in {\cal T}$, and for each $(\mu,s')\in X$, 
there is a node $n'$ child of $n$ labeled by  $s'$, and the 
arc from $n$ to $n'$ is labeled by
$\mu$ and $\prob(\mu,s')$. We will denote by ${\it nodes}(M)$
the set of nodes in  ${\it tree}(M)$, and by ${\it state}(n)$
the state labeling a node $n$.
Example: Figure~\ref{fig:Tree} represents the tree obtained from 
  the probabilistic automaton $M$ of Figure~\ref{fig:PA}.

An {\it adversary} for $M$ is a function $\zeta$ 
that associates to each node $n$ of
${\it tree}(M)$ a transition group among those which are allowed 
in ${\it state}(n)$. More formally, $\zeta : {\it nodes}(M) \rightarrow 
{\it Prob}(A \times S)$ such that $\zeta(n) = (X, \prob)$ implies 
$({\it state}(n), (X, \prob))\in{\cal T}$.

\begin{figure}[ht]
\begin{center}
\epsfbox{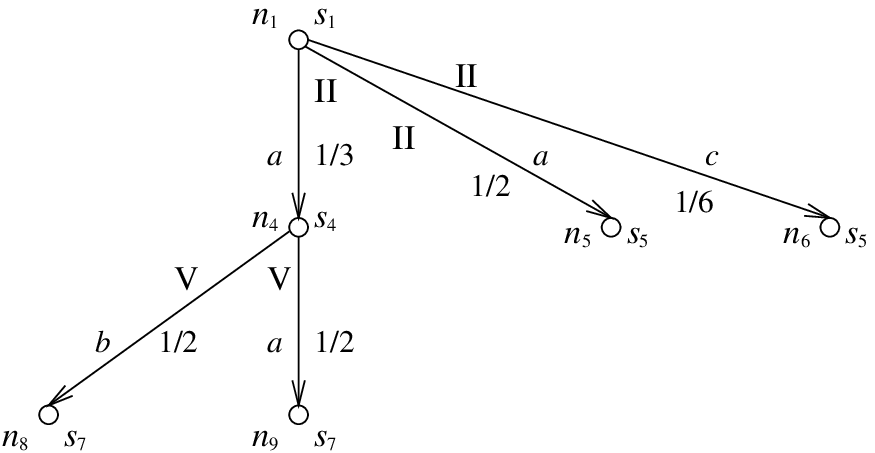}
\caption{etree(M, $\zeta$), where $M$ is the  
probabilistic automaton $M$ of Figure~\ref{fig:PA}, and (the significant part of)
$\zeta$ is defined by $\zeta(n_1) = \mbox{II}$, $\zeta(n_4)=\mbox{V}$}
\label{fig:ETree}
\end{center}
\end{figure}

The {\it execution tree} of an automaton $M=(S,{\cal T},s_0)$ under an adversary $\zeta$,
 denoted by 
${\it etree}(M,\zeta)$, is 
the tree obtained from ${\it tree}(M)$ by pruning all the 
arcs corresponding to transitions which are not in the group selected
by $\zeta$. More formally, ${\it etree}(M,\zeta)$ is a
fully probabilistic 
automaton $(S',{\cal T}',n_0)$, where $S'\subseteq {\it nodes}(M)$, 
$n_0$ is the root of ${\it tree}(M)$, and $(n,(X', \prob'))\in {\cal T}'$ iff
$X' = \{ (\mu,n') \;|\; (\mu, {\it state}(n')) \in X\}$  and $\prob'(\mu, n') =
\prob(\mu, {\it state}(n'))$, where
$(X,\prob) = \zeta(n)$. 
Example: Figure~\ref{fig:ETree}
represents the execution tree of the automaton $M$ of Figure~\ref{fig:PA}, under 
an adversary $\zeta$.

An {\em execution fragment} $\xi$ is any path (finite or infinite) 
from the root of ${\it etree}(M,\zeta)$. The notation $\xi\leq\xi'$ 
means that $\xi$ is a prefix of $\xi'$.
If $\xi$ is  $n_0 \catarrow{p_0}{\mu_0} n_1 \catarrow{p_1}{\mu_1} n_2 
\catarrow{p_2}{\mu_2}
\ldots$, the {\it probability} of $\xi$ is defined as 
$\prob(\xi)=\prod_i p_i$. 
If $\xi$ is maximal, then it is called {\em execution}. 
We denote by ${\it exec}(M,\zeta)$ the set of all executions in ${\it etree}(M,\zeta)$.

We define now a probability on certain sets of executions, following a 
standard construction of Measure Theory. 
Given an execution fragment $\xi$, let $C_\xi = \{ \xi' \in {\it exec}(M,\zeta) \;\mid\;
\xi\leq \xi'\}$ ({\em cone} with prefix $\xi$). Define $\prob(C_\xi)=\prob(\xi)$.
%Let ${\it cones}(M,\zeta)$ the set of all cones. 
Let $\{C_i\}_{i\in I}$ be a countable set of disjoint cones (i.e. $I$ is countable,  
%$\forall i\in I C_i \in {\it cones}(M,\zeta)$ 
and 
$\forall i,j. \;i\neq j \Rightarrow C_i \cap C_j=\emptyset$). 
Then define $\prob(\bigcup_{i\in I}C_i) = \sum_{i\in I}\prob(C_i)$. 
It is possible to show that $\prob$ is well defined, i.e. two 
countable sets of 
disjoint cones with the same union produce the same result for $\prob$.
We can also define the probablity of an empty set of executions as $0$, 
and the probability of the complement of a certain set of executions
as the complement  wrt $1$ of the probability of the set. 
The closure of the cones wrt the empty set, the countable union, and 
the complementation generates what in 
Measure Theory is known as a $\sigma$-field.

\section{The probabilistic asynchronous $\pi$-calculus}
In this section we introduce the probabilistic asynchronous 
$\pi$-calculus ($\pi_{pa}$-calculus for short)
and we give its operational semantics in terms 
of probabilistic automata. 

The $\pi_{pa}$-calculus is obtained from the
asynchronous $\pi$-calculus by replacing 
$\sum_i\nooutpref_i.P_i$ with the following 
{\em probabilistic\ choice\ operator\ }
\[
\sum_i p_i\nooutpref_i.P_i
\]
where the $p_i$'s represents positive probabilities, i.e. they satisfy
$p_i\in(0,1]$
 and $\sum_i p_i = 1$, and the $\alpha_i$'s are input or silent prefixes. 

In order to give the formal definition of the probabilistic model
for $\pi_{pa}$, we find it convenient 
to introduce the following notation for representing transition groups:
given a probabilistic automaton $(S,{\cal T},s_0)$ and 
$s\in S$, we write
\[
s \;\{\catarrow{p_i}{\mu_i}s_i\;|\;i\in I\}
\]
iff 
$(s,(\{(\mu_i,s_i)\;|\;i\in I\},\prob))\in {\cal T}$ and
$\forall i\in I\;p_i = \prob(\mu_i,s_i)$, where $I$ is an index set.
When $I$ is not relevant, we will use the simpler notation
$s \;\{\catarrow{p_i}{\mu_i}s_i\}_i$.
We will also use the notation $s \;\{\catarrow{p_i}{\mu_i}s_i\}_{i:\phi(i)}$,
where $\phi(i)$ is a logical formula depending on $i$, for the set
$s \;\{\catarrow{p_i}{\mu_i}s_i\;|\;i\in I \mbox{ and } \phi(i)\}$.

The operational semantics of a $\pi_{pa}$ process $P$
is defined as a probabilistic automaton whose 
states are the processes reachable from $P$ and the 
${\cal T}$ relation is defined by the rules in Table~\ref{PTS}.
In order to keep the presentation simple, we impose some restrictions on
the syntax of terms (see the caption of Table~\ref{PTS}). In Appendix A
we give an equivalent definition of the operational semantics without these
restrictions. 

\begin{table} 
\begin{center}
\begin{tabular}{|@{\ \ \ }ll@{\ \ }|}  
\hline
\mbox{}&\mbox{}
\\
{\sc Sum}
  &$\sum_i p_i \nooutpref_i.P_i \;\{\catarrow{p_i}{\nooutpref_i} P_i\}_i$
\\
&\mbox{}
\\
{\sc Out}
  &$\bar{x}y \;\{\catarrow{1}{\bar{x}y}{\bf 0}\}$
\\
&\mbox{}
\\
{\sc Open}
 &$\indrule{P\;\{\catarrow{1}{\bar{x}y}P'\}}
           {\nu y P\;\{\catarrow{1}{\bar{x}(y)} P'\}}$
  \ \ \ \  $x\neq y$
\\
&\mbox{}
\\
{\sc Res}
 &$\indrule{P \;\{\catarrow{p_i}{\mu_i} P_i\}_i }
           {\nu y P\;\{\catarrow{p'_i}{\mu_i} \nu y P_i\}_{i: y\not\in 
                                                        {\it fn}(\mu_i)}}$
  \ \ \ \ $\begin{array}{l}
          \exists i.\; y\not\in {\it fn}(\mu_i) 
          \mbox{ and}\\
          \forall i.\; p'_i = p_i/\sum_{j: y\not\in {\it fn}(\mu_j)}p_j
          \end{array}$
\\
&\mbox{}
\\
{\sc Par}
 &$\indrule{P\;\{\catarrow{p_i}{\mu_i} P_i \}_i }
           {P \;|\; Q\;\{\catarrow{p_i}{\mu_i} P_i \;|\; Q \}_i }$
\\
&\mbox{}
\\
{\sc Com}
 &$\indrule{P\;\{\catarrow{1}{\bar{x}y}P'\}\ \ \ \ \ \ 
            Q\;\{\catarrow{p_i}{\mu_i} Q_i\}_i}
 {P \;|\; Q\;\{\catarrow{p_i}{\tau} P' \;|\; Q_i[y/z_i]\}_{i: \mu_i = x(z_i)}
 \cup \{\catarrow{p_i}{\mu_i} P \;|\; Q_i\}_{i: \mu_i \neq x(z_i)}}$
\\
&\mbox{}
\\
{\sc Close}
 &$\indrule{P\;\{\catarrow{1}{\bar{x}(y)}P'\}\ \ \ \ \ \ 
           Q\;\{\catarrow{p_i}{\mu_i} Q_i\}_i}
  {P \;|\; Q\;\{\catarrow{p_i}{\tau} \nu y(P' \;|\; Q_i[y/z_i])\}_{i: \mu_i = x(z_i)}
 \cup \{\catarrow{p_i}{\mu_i} P \;|\; Q_i\}_{i: \mu_i \neq x(z_i)}}$
\\
&\mbox{}
\\
{\sc Cong}
 &$\indrule{P\equiv P' \ \ \ \ \ \ 
   P'\;\{\catarrow{p_i}{\mu_i} Q'_i\}_i \ \ \ \ \ \ \forall i.\; Q'_i\equiv Q_i}
  {P\;\{\catarrow{p_i}{\mu_i} Q_i\}_i }$
\\
&\mbox{} 
\\
\hline
\end{tabular}
\caption{The late-instantiation 
probabilistic transition system of the $\pi_{pa}$-calculus.
In {\sc Sum} we assume that all branches are different, namely,  
if $i\neq j$, then either $\alpha_i\neq \alpha_j$, or $P_i\not \equiv P_j$. 
Furthermore, in {\sc Res} and {\sc Par} we 
assume that all bound variables are distinct from each other, and 
from the free variables. 
}
\label{PTS}
\end{center}
\end{table}

The {\sc Sum} rule models the behavior of a choice process. 
Note that all possible transitions belong to the same group,
meaning that 
the transition is chosen probabilistically by the process itself. 
{\sc Res} models restriction on channel $y$: only the actions 
on channels different from $y$ can be performed and possibly synchronize 
with an external process. 
The probability is redistributed among these actions.
{\sc Par} represents the interleaving of parallel processes. 
All the transitions of the processes involved 
are made possible, and they are kept separated in the 
orininal groups. In this way we model the fact that the 
selection of the process for the next computation step 
is determined by a scheduler. 
In fact, choosing a group corresponds to choosing a process.
{\sc Com}  models communication by handshaking. 
The output action synchronizes with all matching input actions of a partner, 
with the same probability of the input action. The other possible transitions of the 
partner are kept with the original probability as well.  
{\sc Close}  is analogous to {\sc Com}, the only difference is that 
the name being transmitted is private to the sender. 
{\sc Open}  works in combination with {\sc Close} like in the standard
(asynchronous) $\pi$-calculus.
The other rules, {\sc Out} and {\sc Cong}, should be self-explanatory.

\begin{example}\label{example_pi_1}
Consider the  processes $P = {\it rec}_X (1/2\;x(y).{\bf 0} + 1/2\;\tau.X)$, $Q = \bar{x}y$ and 
define $R= P\;|\;Q$. 
The transition groups starting from $R$ are:
\[
R\;\{\catarrow{1/2}{x(y)} Q \;, \;\catarrow{1/2}{\tau} R\} \;\; \;\; \;\; 
R\;\{\catarrow{1/2}{\tau} {\bf 0} \;,\; \catarrow{1/2}{\tau} R\} \;\; \;\; \;\; 
R\;\{\catarrow{1}{\bar{x}y} P\} 
\]
Figure~\ref{fig:ex3} illustrates the probabilistic automaton 
corresponding to $R$. The above transition groups are labeled by I, II and III respectively.
\begin{figure}[ht]
\begin{center}
\epsfbox{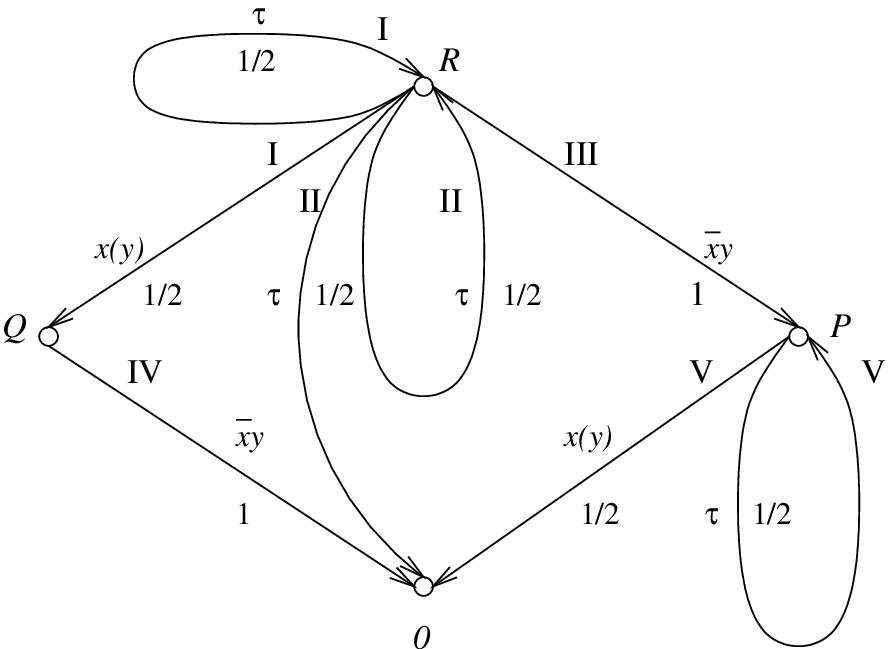}
\caption{The probabilistic automaton of Example~\ref{example_pi_1}}
\label{fig:ex3}
\end{center}
\end{figure}
\end{example}

\begin{example}\label{example_pi_2}
Consider the processes $P$ and $Q$ of
example~\ref{example_pi_1} and define $R= (\nu x)(P\;|\;Q)$. 
In this case the transition groups starting from $R$ are:
\[
R\;\{\catarrow{1}{\tau} R\}  \;\; \;\; \;\; 
R\;\{\catarrow{1/2}{\tau} {\bf 0} \;,\; \catarrow{1/2}{\tau} R\}  
\]
Figure~\ref{fig:ex4} illustrates the probabilistic automaton 
corresponding to this new definition of $R$. The above transition groups are labeled by I and II respectively.
\begin{figure}[ht]
\begin{center}
\epsfbox{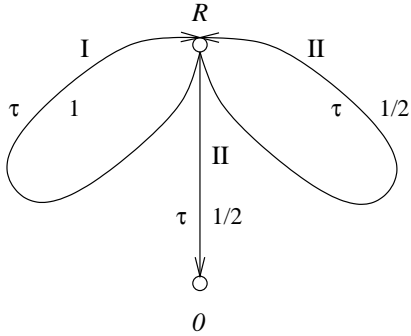}
\caption{The probabilistic automaton of Example~\ref{example_pi_2}\label{fig:ex4}}
\end{center}
\end{figure}
\end{example}

Next example shows that the expansion law does not hold in $\pi_{pa}$. 
This should be no surprise, since the choices associated to the parallel 
operator and to the sum, in $\pi_{pa}$, have a different nature:
the parallel operator gives rise to 
nondeterministic choices of the scheduler, while the sum gives rise
to probabilistic choices of the process.

\begin{example}\label{example_pi_3}
Consider the processes $R_1 = x(z).P\; | \;y(z).Q$ and 
$R_2 = p\; x(z).(P \;|\; y(z).Q) + (1-p)\; y(z).(x(z).P \;| \;Q)$.
The transition groups starting from $R_1$ are:
\[
R_1\;\{\catarrow{1}{x(z)} P \;|\; y(z).Q\}  \;\; \;\; \;\; 
R_1\;\{\catarrow{1}{y(z)} x(z).P \;|\; Q \}  
\]
On the other hand, there is only one transition group starting from 
$R_2$, namely: 
\[
R_2\;\{\catarrow{p}{x(z)} P \;|\; y(z).Q \;\;,\; \;\catarrow{1-p}{y(z)} x(z).P \;|\; Q \}  
\]
%Figure~\ref{fig:ex2_3} illustrates the probabilistic automata 
%corresponding to $R_1$ and $R_2$.
%\begin{figure}[ht]
%\begin{center}
%\epsfbox{Figures/ex2_3.eps}
%\caption{The probabilistic automata $R_1$ and $R_2$ of Example~\ref{example_pi_3}}
%\label{fig:ex2_3}
%\end{center}
%\end{figure}
\end{example}

As announced in the introduction, the parallel operator is associative. 
This property can be easily shown by case analysis. 

\begin{proposition}
For every process $P$, $Q$ and $R$, the probabilistic automata 
of $P\mid(Q\mid R)$ and of $(P\mid Q) \mid R$ are isomorphic, in the sense
that they differ only for the name of the states (i.e. the syntactic structure of the
processes).
\end{proposition}

We conclude this section with a discussion about the design choices of $\pi_{pa}$.

\subsection{The rationale behind the design of $\pi_{pa}$}\label{discussion}

In defining the rules of the operational semantics of $\pi_{pa}$ we felt there was 
only one natural choice, with the exception of the rules {\sc Com} and {\sc Close}. For them  
we could have given a different definition, with respect to which 
the parallel operator would still be associative.

The alternative definition we had considered for {\sc Com} was:
\[
\begin{array}{lll}
{\sc Com}'
 &\indrule{P\;\{\catarrow{1}{\bar{x}y}P'\}\ \ \ \ \ \ 
            Q\;\{\catarrow{p_i}{\mu_i} Q_i\}_i}
 {P \;|\; Q\;\{\catarrow{p'_i}{\tau} P' \;|\; Q_i\}_{i: \mu_i = x(y)}}
 & \begin{array}{l}
          \exists i.\; \mu_i = x(y)  
          \mbox{ and}\\
          \forall i.\; p'_i = p_i/\sum_{j: \mu_j = x(y)}p_j 
          \end{array}
\end{array}
\]
and similarly for {\sc Close}. 

The difference between {\sc Com} and {\sc Com}$'$ is that the latter forces 
the process performing 
the input action  ($Q$) to perform only those actions 
that are compatible with the output action of the partner ($P$). 

At first 
{\sc Com}$'$ seemed to be a reasonable rule. 
At a deeper analysis, however, we discovered that {\sc Com}$'$ imposes certain 
restrictions on the schedulers that, in a distributed setting, would be
rather unnatural. In fact, the natural way of 
implementing the $\pi_a$ communication in a distributed setting
is by representing the input and the output partners as 
processes sharing a common channel. 
When  the sender wishes to communicate, it
puts a message in the channel. When the receiver
wishes to communicate, it tests the channel to see if there is a message, 
and, in the positive case, it retrieves it. 
In case the receiver has a choice guarded by input actions on different 
channels, the scheduler can influence this choice by activating
certain senders instead of others. 
However, if more than one sender has been activated, i.e. more than one channel 
contains data at the moment in which the receiver is activated, then it will be the
receiver which decides internally which channel to select. 
{\sc Com} models exactly this situation. Note that the scheduler
can influence the choices of the receiver by 
selecting certain  outputs to be premises in   
{\sc Com}, and delaying the others by using {\sc Par}. 

With {\sc Com}$'$, on the other hand, when an input-guarded choice is executed, 
the choice of the channel is determined by the scheduler. Thus {\sc Com}$'$ models the 
assumption that the scheduler can only activate (at most) one sender before 
the next activation of a receiver. 

The following example illustrates the difference between {\sc Com}
and {\sc Com}$'$.

\begin{example}
Consider the  processes $P_1 = \bar{x}_1 y$, $P_2 = \bar{x}_2 z$, 
$Q = 1/3\; x_1(y).Q_1 + 2/3 \; x_2(y).Q_2$, and define 
$R= (\nu x_1)(\nu x_2)(P_1\;|\;P_2\;|\;Q)$. 
Under {\sc Com}, the transition groups starting from $R$ are
\[
 R\;\{\catarrow{1/3}{\tau} R_1 , \catarrow{2/3}{\tau} R_2\} \;\; \;\; \;\; 
 R\;\{\catarrow{1}{\tau} R_1\} \;\; \;\; \;\; R\;\{\catarrow{1}{\tau} R_2\}
\]
where $R_1 = (\nu x_1)(\nu x_2)(P_2\;|\;Q_1)$ and 
$R_2 = (\nu x_1)(\nu x_2)(P_1\;|\;Q_2)$.
The first group corresponds to the possibility that 
both $\bar{x}_1$ and $\bar{x}_2$ are available for 
input when $Q$ is scheduled for execution. The other groups correspond
to the availability of only $\bar{x}_1$ and only $\bar{x}_2$
respectively. 

Under {\sc Com}$'$, on the other hand, 
the only possible transition groups are
\[
R\;\{\catarrow{1}{\tau} R_1\}\;\; \;\; \;\; R\;\{\catarrow{1}{\tau} R_2\}
\]
Note that, in both cases, the only possible transitions are those
labeled with $\tau$, because
$\bar{x}_1$ and $\bar{x}_2$ are restricted at the top level.
\end{example}

\section{Solving the electoral problem in $\pi_{pa}$}

\begin{figure}[ht]
\begin{center}
\epsfbox{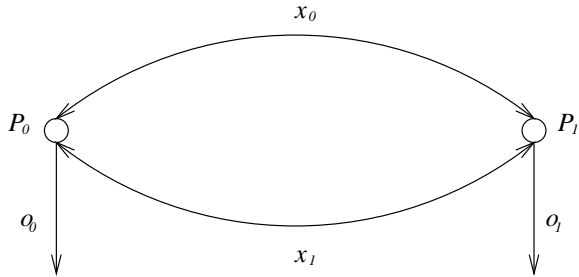} 
\caption{A symmetric network P = $\nu x_0\;\nu x_1 (P_0\;|\;P_1)$. 
The restriction on $x_0$, $x_1$ is made in order to enforce synchronization.}
\label{fig:SNet}
\end{center}
\end{figure}

In \cite{Palamidessi:97:POPL} it has been proved that, 
in certain networks, it is not possible
to solve the leader election problem 
by using  the asynchronous $\pi$-calculus. 
The problem consists in ensuring that 
all processes will reach an agreement (elect a leader) 
in finite time.
One example of such network is the system 
consisting of two symmetric nodes $P_0$ and $P_1$ 
connected by two internal channels $x_0$ and $x_1$ (see Figure~\ref{fig:SNet}).

In this section we will show that it is possible to solve 
the leader election problem for the above network by using  
the $\pi_{pa}$-calculus. 
Following \cite{Palamidessi:97:POPL}, we will assume that 
the processes communicate their decision to the ``external word'' 
by using channels $o_0$ and $o_1$. 

The reason why this problem cannot be solved with the asynchronous 
$\pi$-calculus is that a network with a leader is not symmetric, 
and the asynchronous $\pi$-calculus is not able to force the 
initial symmetry to break. Suppose for example that $P_0$ would
elect itself as the leader after performing a certain sequence of 
actions. By symmetry, and because of lack of synchronous communication,
the same actions may be performed by $P_1$. Therefore $P_1$ 
would elect itself as leader, which means that no agreement 
has been reached.

We propose a solution based on the idea of breaking the symmetry by repeating 
again and again certain random choices, until this goal has been achieved. 
The difficult point is to ensure that it will be achieved with probability $1$
{\it under every possible scheduler}.

Our algorithm works as follows. Each process performs an output on its channel and, 
in parallel, tries to perform an input on both channels. If it succeeds, then it declares 
itself to be the leader. If none of the processes succeeds, it is because both of them 
perform exactly one input (thus reciprocally preventing the other from performing 
the second input). This might occur because the inputs can
be performed only sequentially\footnote{In the $\pi_{pa}$-calculi and in most process algebra 
there is no primitive for simultaneous input action. 
Nestmann has proposed in \cite{Nestmann:98:EXPRESS} the addition of such construct 
as a way of  enhancing the expressive power of the asynchronous 
$\pi$-calculus. Clearly, with this addition, the solution to the 
electoral problem would be immediate.}. In this case, the processes have to try again.
The algorithm is illustrated in Table~\ref{sol_electoral_pb}.

\begin{table}
\begin{center}
\begin{tabular}{|@{\ \ \ }rcl@{\ \ \ }|}  
\hline
&\mbox{}&\mbox{}
\\
$P_i$ & $=$ & \begin{array}[t]{l}
           \bar{x}_i\langle t \rangle \\
           |\; rec_X( \begin{array}[t]{l}
  1/2 \;\tau.x_{i}(b).\begin{array}[t]{l}
              {\it if} \; b \\
              {\it then}\;( \begin{array}[t]{l}
         (1-\varepsilon) \;x_{i \oplus 1}(b).(\bar{o}_i \langle i \rangle \;|\; \bar{x}_i\langle f \rangle) \\
         +  \\
         \varepsilon \; \tau.(\bar{x}_i\langle t \rangle \;|\; X))
                   \end{array} \\
              {\it else} \; \bar{o}_i \langle i \oplus 1\rangle \;)
\end{array}
       \\
   +   \\
   1/2 \;\tau.x_{i \oplus 1}(b).\begin{array}[t]{l}
              {\it if} \; b \\
              {\it then}\;( \begin{array}[t]{l}
         (1-\varepsilon) \;x_i(b).(\bar{o}_i \langle i \rangle \;|\; 
\bar{x}_{i \oplus 1}\langle f \rangle) \\
         +  \\
         \varepsilon \; \tau.(\bar{x}_{i \oplus 1}\langle t \rangle \;|\; X))
                   \end{array} \\
              {\it else} \; \bar{o}_{i} \langle i \oplus 1\rangle \;)
\end{array}
\end{array} 
\end{array}
\\
&\mbox{}&\mbox{}
\\
\hline
\end{tabular}
\caption{A $\pi_{pa}$ solution for the electoral problem in the symmetric 
network
of Figure~\ref{fig:SNet}. Here $i\in \{0, 1\}$ and $\oplus$ is the sum 
modulo 2. }
\label{sol_electoral_pb}
\end{center}
\end{table}

In the algoritm, the selection of the first input is controlled  
by each process with a probabilistic blind choice, 
i.e. a choice whose branches are prefixed by a silent ($\tau$)
action. 
This means that the process commits to the choice of the channel
{\it before} knowing whether it is available. 
It can be proved that this commitment is
essential for ensuring that the leader will be elected
with probability $1$
under every possible adversary scheduler.
The distribution  of the probabilities, 
on the contrary, is not essential. 
This distribution however
affects the efficiency 
(i.e. how soon the synchronization protocol  
converges). 
It can be shown that it is better to split the probability as evenly 
as possible (hence $1/2$ and $1/2$). 

After the first input is performed, a process tries to perform the second 
input. What we would need at this point is a {\it priority choice}, 
i.e. a construct that selects the first branch if the prefix is enabled, and 
selectes the second branch otherwise. With this construct the process 
would perform the input on the other channel when it is available, and 
backtrack to the initial situation otherwise.
Since such construct does not exists
in the $\pi$-calculi, we use probabilities as a way of approximating it. 
Thus we do not guarantee that the first branch will be selected for sure 
when the prefix is enabled, but we guarantee that it will be selected with 
probability close to $1$: the symbol $\varepsilon$ represents a very small 
positive number. Of course, the smallest $\epsilon$ is, the more efficient 
the algorithm is.

When a process, say $P_0$, succeeds to perform both inputs, then it declares 
itself to be the leader.
It also notifies this decision to the other process.
For the notification we could use a different channel, or we may use the same 
channel, 
provided that we have a way to communicate that the output on such channel 
has now 
a different meaning. We follow this second approach, and we use boolean values 
 {\bf t} and {\bf f} for messages. We stipulate that
 {\bf t} means that the leader has not been decided yet,
 while {\bf f} means that it has been decided. 
Notice that the symmetry is broken exactly when one process succeeds 
in performing both inputs.
 
In the algorithm we make use of the if-then-else construct, 
which is defined by the structural rules
\[
{\it if\ {\bf t}\ then\ P \ else\  Q} \equiv P \;\;\;\;\;\;\;\; 
{\it if\ {\bf f}\ then\ P \ else\  Q} \equiv Q
\] 
As discussed in \cite{Nestmann:96:CONCUR}, these features 
(booleans and if-then-else) can be translated into the asynchronous 
$\pi$-calculus, and therefore in $\pi_{pa}$. 

\subsection*{Correctness of the algorithm}
 
We prove now that the algorithm 
is correct, namely that the probability that 
a leader is eventually elected is $1$ under every scheduler.

In the following we use pairs to denote the $\tau$ transitions
corresponding to the execution of 
the blind choice. A pair $(i,j)$ will mean that process 
$i$ has selected channel $j$. We will call such transitions {\it random draws}.

\begin{definition}
A sequence $d_1,d_2,\ldots,d_n,\ldots$ of random draws is {\it alternated} 
iff $\forall \; k$, if 
$d_k=(i,j)$ then $d_{k+1}=(i,j)$ or $d_{k+1}=(i \oplus 1,j \oplus 1)$.  
\end{definition}

Note that a sequence is alternated iff for every two draws $(i,j),(i',j')$ 
if $i=i'$ then $j=j'$.
 
For the proof, we are going to consider, at first, a modified algorithm 
where the inner choice ($(1-\varepsilon)\ldots + \varepsilon\ldots$) 
is replaced by a priority choice.

\begin{lemma}\label{electoral1}
Consider an execution fragment $\xi$
of the process $\nu x_0\;\nu x_1 (P_0\;|\;P_1)$ and the algorithm 
of Table~\ref{sol_electoral_pb} modified by using the 
priority choice. Let $d_1,d_2,\ldots,d_n$ be the 
sequence of random draws in $\xi$. 
Assume that, for some $k<n$, 
$d_k=(i,j)$, $d_{k+1}= \ldots = d_{n-1} = (i \oplus 1, j \oplus 1)$, 
and $d_n=(i \oplus 1,j)$.  
Then, under every adversary, all the executions in the cone of $\xi$ 
terminate with the election of a leader, 
and they contain no more random draws.
\end{lemma}

\noindent{\bf Proof} (Sketch)
If at a certain point both processes have committed to the same channel, then
only one of them will be able to perform the input action on that channel, 
whereas the other one is blocked waiting
to perform an input action on the same channel. 
The process that is able to make
the input action will therefore be able to 
make the second input action too and 
will become the leader. The other process 
will finally be enabled to make the input on the 
channel on which it was blocked, and will 
receive the notification 
that the other 
process has become the leader. 
Neither processes select the recursive branch and therefore
no more random draws are made.
\qed

\begin{lemma}\label{electoral3}
The probability that a sequence of random draws of length 
$n$ is alternated is $1/2^{n-1}$.
\end{lemma}
\noindent{\bf Proof} Obvious, by induction on $n$, 
and by the observation that the random draws
are independent. 
\qed

\begin{proposition}\label{electoral4}
Consider the process $\nu x_0\;\nu x_1 (P_0\;|\;P_1)$
and the algorithm of Table~\ref{sol_electoral_pb} modified by using the 
priority choice.
The probability of the executions which  contain (at least) $n$ random draws, 
for $n\geq 2$, is at most $1/2^{n-2}$ 
under every adversary.
\end{proposition}

\noindent{\bf Proof} 
By Lemma~\ref{electoral1} the first $n-1$ random draws must be alternated
(otherwise the leader would have been elected earlier). 
By Lemma~\ref{electoral3} such alternated  sequence has probability $1/2^{n-2}$. 
Note that the maximum 
probability $1/2^{n-2}$ corresponds to the worst possible case of an adversary 
which tries to delay the election of a leader as much as possible, by scheduling the 
processes in such a way that a process tries to perform  the second input
only when the channel is not available.
\qed
\ \\

We are now ready to prove the correctness of our algorithm. 

\begin{proposition}\label{electoral5}
Consider the process $\nu x_0\;\nu x_1 (P_0\;|\;P_1)$
and the algorithm of table \ref{sol_electoral_pb} (with no modifications).
The probability of the executions which  contain (at least) 
$n$ random draws, for $n\geq 2$, is at most 
\[
\frac{(1+\varepsilon)^{n-2}}{2^{n-2}}
\] 
under every adversary.
\end{proposition}

\noindent{\bf Proof} 
The proof proceeds like in the proof of Proposition~\ref{electoral4}, 
with the exception that we need to consider also the possibility that 
a leader is {\it not} elected after a draw which breaks the alternation. 
Such event occurs with probability $\varepsilon$. 
The probability that an execution contains $n-1$ draws 
where the alternation is violated $k$ times is therefore
\[
\frac{1}{2^{n-2}}\;\frac{ (n-2)!}{k! (n-2-k)!}\;\varepsilon^k
\]
The sum of these probability for all possible values of $k$ 
is 
\[
\frac{1}{2^{n-2}}
\sum_{k=0}^{n-2}
\frac{(n-2)!}{k!(n-2-k)!}\;\varepsilon^k \;=\;
\frac{(1+\varepsilon)^{n-2}}{2^{n-2}} 
\]
\qed
\ \\

As a consequence of this proposition we finally obtain the 
correctness of our algorithm:  
\begin{theorem}
Consider the process $\nu x_0\;\nu x_1 (P_0\;|\;P_1)$
and the algorithm of table \ref{sol_electoral_pb}   (without modifications).
The probability that the leader
 is eventually elected
is $1$
under every adversary.
\end{theorem}

\noindent{\bf Proof} 
An execution does not elect a leader only if it is 
infinite and contains an infinite number of random draws. 
By Proposition~\ref{electoral5} the probability of
the execution fragments which contain at least $n$ random draws is
at most
\[
\frac{(1+\varepsilon)^{n-2}}{2^{n-2}} 
\]
ince $\varepsilon<1$, this probability converges to $0$ for 
$n\rightarrow\infty$.
\qed
\ \\

We conclude this section with the observation that,
 if we modify the blind choice 
to be a choice prefixed with the input actions 
which come  immediately afterward, then 
the above theorem would not hold anymore. 
In fact, we can define a scheduler which selects the processes 
in alternation, and
which suspends a process, and  activates the other, 
immediately after the first has made a random choice and performed an input. 
The
latter will be forced (because of the guarded choice) to perform 
the input on the other channel. Then the scheduler will proceed with the first 
process, which at this point can only backtrack. 
Then it will schedule the second process again, which will 
also be forced to backtrack, and so on. 
Since all the choices of the processes are obligated in this scheme, the
scheduler will produce an infinite (unsuccessful) execution with probability 
$1$.

\section{Implementation of $\pi_{pa}$ in a Java-like language}
In this section we propose an implementation of the 
{\it synchronization}-{\it closed}
$\pi_{pa}$-calculus, namely the subset of $\pi_{pa}$ 
consisting of processes in which all occurrences of 
communication actions $x(y)$ and $\bar{x}y$ are under 
the scope of a restriction operator $\nu x$. 
This means that all communication actions 
are forced to synchronize. 

The implementation is written in a Java-like language
following the idea outlined in Section~\ref{discussion}. 
It is compositional wrt all the operators, and distributed, 
i.e. homomorphic wrt the parallel operator.

Channels are implemented as one-position buffers, 
namely as objects of the following class:
\begin{verbatim}
     class Channel {
             Channel message;
             boolean isEmpty;

             public void Channel() {
                     isEmpty = true;
             }

             public synchronized void send(Channel y) {
                     while (!isEmpty)  wait();
                     isEmpty = false;
                     message = y;
                     notifyAll();
             }

             public synchronized GuardState test_and_receive() {
                     GuardState s = new GuardState();
                     if (! isEmpty) { s.test = true;
                                      s.value = message;
                                      isEmpty = true;
                                      return s; }
                     else { s.test = false;
                            s.value = null;
                            return s; }	 
              }
     }

     class GuardState {
             public boolean test;
             public Channel value;
     }
\end{verbatim}

The methods \verb|send| and  \verb|test_and_receive| are 
used for implementing the output and the input actions respectively. 
They are both {\it synchronized}, because the test for the 
emptyness (resp. non-emptyness) of the channel, and the subsequent 
placement (resp. removal) of a datum, must be done atomically.

Note that, in principle, the receive method could have been defined 
dually to the send method, i.e. read and remove a datum if present, and 
suspend (wait) otherwise. This definition would work for  
input prefixes which are not in the context of a choice. However, 
it does not work
for input guarded choice. In order to simulate correctly the behavior of  the
input guarded choice, in fact, we should check continuously for input 
events, 
until we find one which is enabled. Suspending when one of 
the input guards is not enabled would be incorrect.
Our definition of \verb|test_and_receive| circumvent this problem 
by reporting a failure to the caller, instead of suspending it.

Given the above representation of  channels, the $\pi_{pa}$-calculus can be 
implemented by using the following encoding $\ls\cdot\rs$:

\subsubsection*{Probabilistic choice}
\[\ls\,  \sum_{i=1}^m p_i x_i(y).P_i + \sum_{i=m+1}^n p_i \tau.P_i \,\rs  =\]
\begin{verbatim}
     { boolean choice = false;
       GuardState s = new GuardState();
       float x;
       Random gen = new Random();
       while (!choice) {
          x = 1 - gen.nextFloat();  % nextFloat() returns a real number in [0,1)
\end{verbatim}     
\verb|          if (|$0$\verb| < x <= | $p_1$ \verb|) |\\
\verb|               { s = x1.test_and_receive(); |\\
\verb|                 if (s.test) { y = s.value;|
$\ls\, P_1 \,\rs$\\              
\verb|                               choice = true; }|\\                      
\verb|               }|\\
\verb|          ...|\\ 
%\verb|     |\\
\verb|          if (|$p_1+p_2+...+p_{m-1}$ \verb| < x <= | $p_1+p_2+...+p_m$\verb|)|\\
\verb|               { s = xm.test_and_receive();|\\
\verb|                 if (s.test) { y = s.value;|
$\ls\, P_m \,\rs$\\
\verb|                               choice = true; }|\\                   
\verb|               }|\\
%\verb|     |\\
\verb|          if (|$p_1+p_2+...+p_m$\verb| < x <= |$p_1+p_2+...+ p_{m+1}$\verb|) |\\
\verb|               {|
$\ls\, P_{m+1} \,\rs$\\
\verb|                 choice = true; }|\\ 
\verb|          ...|\\ 
%\verb|     |\\
\verb|          if (|$p_1+p_2+...+p_{n-1}$\verb| < x <= |$p_1+p_2+...+ p_{n}$\verb|) |\\
\verb|               {|
$\ls\, P_n \,\rs$\\
\verb|                 choice = true; }|\\ 
\verb|     }|

Note that with this implementation, when no input guards are enabled, the process 
keeps performing  internal (silent) actions instead of suspending.

\subsubsection*{Output action}
\[\ls\, \bar{x}y \,\rs =\verb+ {x.send(y);} +\] %%%% changed

\subsubsection*{Restriction}
\[\ls\, \nu x P \,\rs =\mbox{ \tt \{Channel x = new Channel(); }\ls\, P \,\rs \mbox{ \tt \} }\]

\subsubsection*{Parallel}
If our language is provided with a parallel operator, then we can just have 
a homomorphic mapping:
\[\ls\, P_1 \;|\; P_2 \,\rs = \ls\, P_1 \,\rs \;|\; \ls\, P_2 \,\rs\]
In Java, however, there is no parallel operator.
In order to mimic it, a possibility is to define 
a new class for each process we wish to compose in parallel, and then create and 
start an object of that class:\\
\verb| |\\
\verb|     class processP1 extends Thread {|\\
\verb|               public void run() {|\\
\verb|                        | 
$\ls\, P_1 \,\rs$\\
\verb|               }|\\
\verb|     }|\\
\[\ls\, P_1 \;|\; P_2 \,\rs = \mbox{ \tt \{ new processP1.start(); }\ls\, P_2 \,\rs\mbox{ \tt \} }\]

\subsubsection*{Recursion}
Remember that the process $ rec_XP $  represents a process $X$ defined as $X  \stackrel{\rm def}{=} P$,
where $P$ may contain occurrences of $X$. For each such process, define the 
following class:\\
\verb| |\\
\verb|     class X {|\\
\verb|           static public void exec() {|\\
\verb|                    | 
$\ls\, P \,\rs$\\
\verb|           }|\\
\verb|     }|\\
\verb| |\\
Then define:
\[\ls\, rec_XP \,\rs = \verb|{ X.exec(); }| \] %%%% changed
\[\ls\, X \,\rs = \verb|{ X.exec(); }| \] %%%% changed

\bibliographystyle{plain}
\bibliography{biblio_cat}

\begin{thebibliography}{10}

\bibitem{Amadio:98:TCS}
Roberto~M. Amadio, Ilaria Castellani, and Davide Sangiorgi.
\newblock On bisimulations for the asynchronous $\pi$-calculus.
\newblock {\em Theoretical Computer Science}, 195(2):291--324, 1998.
\newblock An extended abstract appeared in {\em Proceedings of CONCUR '96},
  LNCS 1119: 147--162.

\bibitem{Boudol:92:REPORT}
G{\'e}rard Boudol.
\newblock Asynchrony and the $\pi$-calculus (note).
\newblock Rapport de Recherche 1702, INRIA, Sophia-Antipolis, 1992.

\bibitem{Honda:91:ECOOP}
Kohei Honda and Mario Tokoro.
\newblock An object calculus for asynchronous communication.
\newblock In Pierre America, editor, {\em Proceedings of the European
  Conference on Object-Oriented Programming (ECOOP)}, volume 512 of {\em
  Lecture Notes in Computer Science}, pages 133--147. Springer-Verlag, 1991.

\bibitem{Milner:99:BOOK}
Robin Milner.
\newblock {\em Communicating and mobile systems: the $\pi$-calculus}.
\newblock Cambridge University Press, 1999.

\bibitem{Milner:92:IC}
Robin Milner, Joachim Parrow, and David Walker.
\newblock A calculus of mobile processes, {I and II}.
\newblock {\em Information and Computation}, 100(1):1--40 \& 41--77, 1992.

\bibitem{Milner:93:TCS}
Robin Milner, Joachim Parrow, and David Walker.
\newblock Modal logics for mobile processes.
\newblock {\em Theoretical Computer Science}, 114(1):149--171, 1993.

\bibitem{Nestmann:98:EXPRESS}
Uwe Nestmann.
\newblock On the expressive power of joint input.
\newblock In Catuscia Palamidessi and Ilaria Castellani, editors, {\em
  {EXPRESS} '98: Expressiveness in Concurrency}, volume 16.2 of {\em Electronic
  Notes in Theoretical Computer Science}. Elsevier Science B.V., 1998.

\bibitem{Nestmann:96:CONCUR}
Uwe Nestmann and Benjamin~C. Pierce.
\newblock Decoding choice encodings.
\newblock In Ugo Montanari and Vladimiro Sassone, editors, {\em Proceedings of
  {CONCUR} '96: Concurrency Theory (7th International Conference, Pisa, Italy,
  August 1996)}, volume 1119 of {\em Lecture Notes in Computer Science}, pages
  179--194. Springer-Verlag, 1996.
\newblock Full version to appear in {\em Information and Computation}.

\bibitem{Palamidessi:97:POPL}
Catuscia Palamidessi.
\newblock Comparing the expressive power of the synchronous and the
  asynchronous {$\pi$}-calculus.
\newblock In {\em Conference Record of {POPL}~'97: The 24th {ACM}
  {SIGPLAN}-{SIGACT} Symposium on Principles of Programming Languages}, pages
  256--265, Paris, France, 1997.

\bibitem{Rabin:94:HOARE}
Michael~O. Rabin and Daniel Lehmann.
\newblock On the advantages of free choice: {A} symmetric and fully distributed
  solution to the dining philosophers problem.
\newblock In A.~W. Roscoe, editor, {\em A Classical Mind: Essays in Honour of
  C.A.R. Hoare}, chapter~20, pages 333--352. Prentice Hall, 1994.
\newblock An extended abstract appeared in the {\em Proceedings of POPL'81},
  pages 133-138.

\bibitem{Segala:95:PhD}
Roberto Segala.
\newblock {\em Modeling and Verification of Randomized Distributed Real-Time
  Systems}.
\newblock PhD thesis, Department of Electrical Engineering and Computer
  Science, Mass\-a\-chu\-setts Insti\-tute of Tech\-no\-logy, June 1995.
\newblock Available as Technical Report MIT/LCS/TR-676.

\bibitem{Segala:95:NJC}
Roberto Segala and Nancy Lynch.
\newblock Probabilistic simulations for probabilistic processes.
\newblock {\em Nordic Journal of Computing}, 2(2):250--273, 1995.
\newblock An extended abstract appeared in {\em Proceedings of CONCUR '94},
  LNCS 836: 22--25.

\bibitem{Glabbeek:95:IC}
Rob~J. van Glabbeek, Scott~A. Smolka, and Bernhard Steffen.
\newblock Reactive, generative and stratified models of probabilistic
  processes.
\newblock {\em Information and Computation}, 121(1):59--80, 1995.

\bibitem{Vardi:85:FOCS}
Moshe~Y. Vardi.
\newblock Automatic verification of probabilistic concurrent finite-state
  programs.
\newblock In {\em Proceedings of the 26th Annual Symposium on Foundations of
  Computer Science}, pages 327--338, Portland, Oregon, 1985. IEEE Computer
  Society Press.

\bibitem{Yi:92:IFIP}
Wang Yi and Kim~G. Larsen.
\newblock Testing probabilistic and nondeterministic processes.
\newblock In {\em Proceedings of the 12th IFIP International Symposium on
  Protocol Specification, Testing and Verification}, Florida, USA, 1992. North
  Holland.

\end{thebibliography}

\section*{Appendix A}
Table~\ref{PTS_Alt} presents an equivalent 
transition system for the $\pi_{pa}$-calculus
where no assumptions on the bound variables are made. 
Note that the side condition on the rule {\sc Sum} 
is necessary for treating cases like $1/2 \;x(y).{\bf 0} + 1/2\; x(y).{\bf 0}$.
This condition 
could be eliminated by assuming that 
the transition groups are multiset instead than sets. 

\begin{table}[ht] 
\begin{center}
\begin{tabular}{|@{\ \ \ }ll@{\ \ }|}  
\hline
\mbox{}&\mbox{}
\\
{\sc Sum}
  &$\sum_i p_i \nooutpref_i.P_i \;\{\catarrow{p'_i}{\nooutpref_i} P_i\}_i$
 \ \ \ \ $p'_i = p_i/\sum_{j:\nooutpref_j={\nooutpref_i} \;
P_j\equiv P^i}p_j$
\\
&\mbox{}
\\
{\sc Out}
  &$\bar{x}y \;\{\catarrow{1}{\bar{x}y}{\bf 0}\}$
\\
&\mbox{}
\\
{\sc Open}
 &$\indrule{P\;\{\catarrow{1}{\bar{x}y}P'\}}
           {\nu y P\;\{\catarrow{1}{\bar{x}(y)} P'\}}$
  \ \ \ \ $x\neq y$
\\
&\mbox{}
\\
{\sc Res}
 &$\indrule{P \;\{\catarrow{p_i}{\mu_i} P_i\;|\; i\in I\} }
           {\nu y P\;\{\catarrow{p'_i}{\mu_i} \nu y P_i \;|\; 
                i\in I \mbox{ and }  y\not\in {\it fn}(\mu_i) \}}$
  \ \ \ \ $\begin{array}{l}
          \exists i\in I.\; y\not\in {\it fn}(\mu_i), \\
          \forall i\in I.\; y\not\in {\it bn}(\mu_i), 
          \mbox{ and}\\
          \forall i\in I.\; p'_i = p_i/\sum_{j: y\not\in {\it fn}(\mu_j)}p_j
          \end{array}$
\\
&\mbox{}
\\
{\sc Par}
 &$\indrule{P\;\{\catarrow{p_i}{\mu_i} P_i \}_i }
           {P \;|\; Q\;\{\catarrow{p_i}{\mu_i} P_i \;|\; Q \}_i }$
  \ \ \ \ $\forall i.\; {\it bn}(\mu_i)\cap{\it fn}(Q) =\emptyset$
\\
&\mbox{}
\\
{\sc Com}
 &$\indrule{P\;\{\catarrow{1}{\bar{x}y}P'\}\ \ \ \ \ \ 
            Q\;\{\catarrow{p_i}{\mu_i} Q_i\;|\;i\in I\}}
 {P \;|\; Q\;\{\catarrow{p_i}{\tau} P' \;|\; Q_i[y/z_i] \;|\;
          i\in I \mbox{ and } \mu_i = x(z_i) \}
 \cup \{\catarrow{p_i}{\mu_i} P \;|\; Q_i \;|\;
          i\in I \mbox{ and } \mu_i \neq x(z_i) \} }$
\\
&\mbox{}
\\
{\sc Close}
 &$\indrule{P\;\{\catarrow{1}{\bar{x}(y)}P'\}\ \ \ \ \ \ 
           Q\;\{\catarrow{p_i}{\mu_i} Q_i\;|\;i\in I\}}
  {P \;|\; Q\;\{\catarrow{p_i}{\tau} \nu y(P' \;|\; Q_i[y/z_i])\;|\;
          i\in I \mbox{ and } \mu_i = x(z_i) \}   
  \cup \{\catarrow{p_i}{\tau} P \;|\; Q_i \;|\;
          i\in I \mbox{ and } \mu_i \neq x(z_i) \} }$
\\
&\mbox{}
\\
{\sc Cong}
 &$\indrule{P\equiv P' \ \ \ \ \ \ 
   P'\;\{\catarrow{p_i}{\mu_i} Q'_i\}_i \ \ \ \ \ \ \forall i.\; Q'_i\equiv Q_i}
  {P\;\{\catarrow{p_i}{\mu_i} Q_i\}_i }$
\\
&\mbox{} 
\\
\hline
\end{tabular}
\caption{Alternative formulation of the 
probabilistic transition system for the $\pi_{pa}$-calculus.}
\label{PTS_Alt}
\end{center}
\end{table}

\end{document}